**Research article**

# Intense ultraviolet perturbations on aquatic primary producers

**Mayrene Guimarais[1], Rolando Cárdenas[2] and J.E. Horvath[3]**

[1] {Marine Ecology Group, Center for Research of Coastal Ecosystems, Cayo Coco, Ciego de Avila, Cuba . Phone 53 33 301104 ext 117 e-mail: mayrene@ciec.fica.inf.cu}

[2] {Department of Physics, Universidad Central de Las Villas, Santa Clara, Cuba. Phone 53 42 281109 Fax 53 42 281130 e-mail: rcardenas@uclv.edu.cu}

[3] {Departmento de Astronomia, Instituto de Astronomia, Geofísica e Ciências Atmosféricas, Universidade de São Paulo, Brazil. Phone (11) 3091 2710 e-mail: foton@astro.iag.usp.br}

## Abstract

During the last decade, the hypothesis that one or more biodiversity drops in the Phanerozoic eon, evident in the geological record, might have been caused by the most powerful kind of stellar explosion so far known (Gamma Ray Bursts) has been discussed in several works. These stellar explosions could have left an imprint in the biological evolution on Earth and in other habitable planets. In this work we calculate the short-term lethality that a GRB would produce in the aquatic primary producers on Earth. This effect on life appears as a result of ultraviolet (UV) re-transmission in the atmosphere of a fraction of the gamma energy, resulting in an intense UV flash capable of penetrating ~ tens of meters in the water column in the ocean. We focus on the action of the UV flash on phytoplankton, as they are the main contributors to global aquatic primary productivity. Our results suggest that the UV flash could cause an hemispheric reduction of phytoplankton biomass in the upper mixed layer of the World Ocean of around 10%, but this figure can reach up to 25 % for radiation-sensitive picoplankton species, and/or in conditions in which DNA repair mechanisms are inhibited.

## 1 Introduction

The ultraviolet shielding problem has attracted the attention of the biologists for a long time. There are many evidences indicating the lack of an atmospheric ultraviolet shield during the Archean eon, and as a consequence the possible onset of a harsh photobiological regime in the planet's surface, listed among the reasons of why continents might not have been conquered by life until the appearance of the ozone layer. Thus, photobiology stands as one of the drivers of biological evolution in our planet. Radiations in general have the dual role of sterilizing non-resistant species and favouring speciation of the surviving ones, due to DNA mutations and expected availability of ecological niches. Therefore, radiation bursts are plausible hypotheses to explain biodiversity drops and its subsequent increases, as in the case of the Cambrian explosion and Phanerozoic extinctions. One of the natural mechanisms capable of delivering on Earth radiation bursts intense enough are stellar explosions, provided the explosion occurs not too far (Thorsett 1995, Scalo and Wheeler 2000, Galante and Horvath 2007; Martin *et al* 2009). These explosions typically occur in very massive stars, progenitors of the so-called gamma ray bursts (GRB's) and associated supernovae. More specifically, it has been suggested that the Ordovician-Silurian mass extinction was caused by a GRB (Melott and Thomas 2009; Melott *et al* 2004) and that the minor marine extinction in the Pleistocene-Pliocene of tropical bivalves was the consequence of a nearby (ordinary?) supernova (Benítez, Maiz-Apellaniz and Canelles 2002). It is worth noting that the last hypothesis is receiving support by isotopic anomalies in marine sediments of the Pleistocene, namely of the same epoch in which our Solar System neared the Scorpius-Centaurus association of massive stars (Fields, Hochmuth and Ellis 2005). Currently, there is at least one star close enough to Earth as to be considered dangerous because of its potential explosion as a supernova and even the emission of a GRB. It is WR 104, located just 8000 light years away from us (Tuthill *et al* 2008), although recent spectroscopic measurements suggest a pitch of the gamma beam ~ 30-40 degrees from Earth, thus probably preventing the incidence on our atmosphere. In general, statistical estimates suggest a very low incidence probability, mainly because of a small solid angle gamma emission (Mézáros 2001), but there might be undetected binary systems close enough to be problematic.

An earlier work (Galante and Horvath 2007) has compiled and discussed the several effects that a stellar explosion can cause on Earth's atmosphere and biosphere. The best studied is the depletion of the ozone layer, allowing more solar UV to reach the planet's ground during several years. In this work, however, we focus on another important short-term effect: the brief and immediate UV-flash reaching the ground as a result of reprocessing the gamma energy in the atmosphere. A closely related phenomenon is also known form solar observations:, arguably our Sun has the potential of sporadic flares intense enough to cause ecological catastrophes, but so far none of them has been recorded and confirmed. However, it is clear that even modest depletions of the ozone layer can significantly influence terrestrial biota, through the enhanced solar UV flux reaching every day the surface of Earth. That is why since the 1970's much attention has been given to the current depletion of the ozone layer, mainly in the context of the anthropogenic global warming. Actually, the ozone hole over Antarctica and a potential future one over Europe have been defined as tipping points of our planet (Lenton *et al.* 2008; IPCC 2007).

In this work we use tools developed to model biological effects of current ozone depletion in order to do some estimates of the *immediate* lethality that a stellar explosion or an unusually intense solar flare would cause on phytoplankton. These are the main primary producers and the starting point of the food web in central ocean basins, and are also important in coastal and freshwater ecosystems. Astrophysical calculations based on star formation rate suggest that in the last few billion years each planet in our galaxy would have been affected by a GRB (Scalo and Wheeler 2004). We thus focus in the short-term lethality that would produce on Earth the so-called "typical" GRB of the last billion years: a burst arriving from 3000-6000 light years away and delivering ~ 100 $kJ/m^2$ of gamma energy at the top of the atmosphere. The main short-term difference between stellar explosions and solar flares will be in the *specific* ultraviolet spectrum deposited at ground, provided total energy is equal. The photobiological methods used to estimate biological damage are, though, the same for each case.

## 2 Materials and methods

### 2.1 - The interaction of stellar gamma radiation with the atmosphere

The interaction of the gamma burst with the atmosphere has been considered and we adopted the results of Martin *et al.* (2009) in this work. In first place, the fraction of gamma photons directly reaching the ground is negligible, because of the large Compton cross-section with electrons from the molecules of the atmosphere. The free electrons would then excite other molecules, causing a rich aurora-like spectrum, which reaches the sea level. The ultraviolet fraction of this spectrum (termed the UV-flash) is a major danger for life (Galante and Horvath 2007). The duration of the UV-flash would be the same of the GRB (around 10 seconds), with a high intensity and even including the very deleterious UV-C band in the wavelength range $260-280nm$. The interactions of these UV flash photons in water and their efficiency for phytoplankton damage is our concern in this work.

### 2.2 - Radiative transfer in water and effective doses

We considered an average ocean albedo of 6.6 % for zenithal angles not greater than 70 degrees, as reported in Cockell (2000). This was employed to calculate the GRB-UV spectrum just below the ocean surface $E_0(\lambda,0^-)$. We used the classification of optical ocean water types originally presented in Jerlov (1951, 1964, 1976). Consequently, we use the attenuation coefficients $K(\lambda)$ of UVR in oceanic water types I, II and III as in (Peñate *et al* 2010). These optical water types can roughly be identified as oligotrophic, mesotrophic and eutrophic, respectively. However, we also included the intermediate types IA and IB. We utilized biological action spectrum for DNA damage $e(\lambda)$ following Cockell (2000). Then, the (effective) biological irradiances or dose rates $E^*(z)$ at depth $z$ follow from :

$$E^*(z) = \sum e(\lambda) E_0(\lambda, 0^-) e^{-K(\lambda)z} d\lambda \qquad (1)$$

The (effective) biological fluences or doses $F^*(z)$ are given by

$$F^*(z) = E^*(z)\Delta t \qquad (2)$$

where $\Delta t$ is the exposure time to UVR.

We also consider that, just before the UV flash, phytoplankton were homogeneously distributed in the upper mixed layer (UML) of the ocean, due to the mixing action of Langmuir currents and related circulation patterns. The depth of UML depends on ocean surface winds and other factors, but after averaging its value for 13 locations (Agustí and Llabrés 2007) we consider it to be 30 meters, quite a typical value.

### 2.3 – The estimation of induced lethality

Experiments with phytoplankton stressed by UVR are typically done exposing them to solar radiation during several hours. This is not a scenario close enough to the one we study, given the low intensity of solar UV, as compared to the GRB UV-flash. Therefore, as done by some of us in (Galante and Horvath 2007), we chose the results of (Gascon *et al* 1995). These authors intensely irradiated a representative set of bacteria with a "hard" wavelength ($\lambda = 254\, nm$) of the UV-C band. We believe that the more radiation-sensitive phytoplankton would behave as *Escherichia coli*, the intermediate as the aquatic bacterium *Rhodobacter sphaeroides* (wild type and phototrophically grown strain), while the toughest would parallel the soil bacterium *Rhizobium meliloti*. We also analyzed the case in which repair mechanisms would be inhibited: very cold waters or a night-time UV-flash (because at night cell division is synchronized in oceanic phytoplankton (Agawin and Agustí 2005), making them much more radiation sensitive). To account for this last scenario we use the data in (Gascon *et al* 1995) for strains in which repair is inhibited due to the lack of *recA* gene. These data are namely available for the two extremes of our "survival band" (*E. coli* and *R. meliloti*). Strains having above gene are denoted *recA*$^{+}$, while the absence is indicated by *recA*$^{-}$.

Starting with the classical model for survival curves of irradiated cells,

$$S = e^{-\alpha F} \qquad (3),$$

where $S$ is the survival fraction, $\alpha$ is the slope and $F$ is the dose or fluence, we have introduced some significant refinements. Since the effective biological dose $F^*$ calculated from eqs. (1) and (2) needs to be employed, we propose a refined survival model:

$$S(z) = e^{-\alpha^* F^*(z)} \qquad (4)$$

where $S(z)$ is the surviving fraction at depth $z$, $\alpha^*$ is the new (effective biological) slope, and $F^*(z)$ is the effective biological dose or fluence at depth $z$.

The slopes $\alpha^*$ are a measure of the radiosensitivity of the species. We determine them considering that the reported doses $F$ in Gascon *et al* (1995) follow the simple formula:

$$F = E\Delta t \qquad (5)$$

Dividing eq. (2) by eq. (5) we obtain:

$$\frac{F^*}{F} = \frac{E^*}{E} \qquad (6)$$

Both $F$ and $E$ are given in Gascon *et al* (1995), while $E^*$ was determined by Cockell (2000) by biologically weighting it, so above equation allows the calculation of $F^*$ for each species. We then obtained the biological effective dose for which 10% of the cells survive ($F^*_{10}$) using the $F_{10}$ values for each species reported in Gascon *et al* (1995), and finally found the new slope $\alpha^*$.

## 3 Results

### 3.1 Radiation transfer and effective doses in the ocean

The set of attenuation coefficients used is shown in Fig. 1. Notice that in the wavelength range used (260-350nm), the optical quality of types I, IA and IB is not very different.

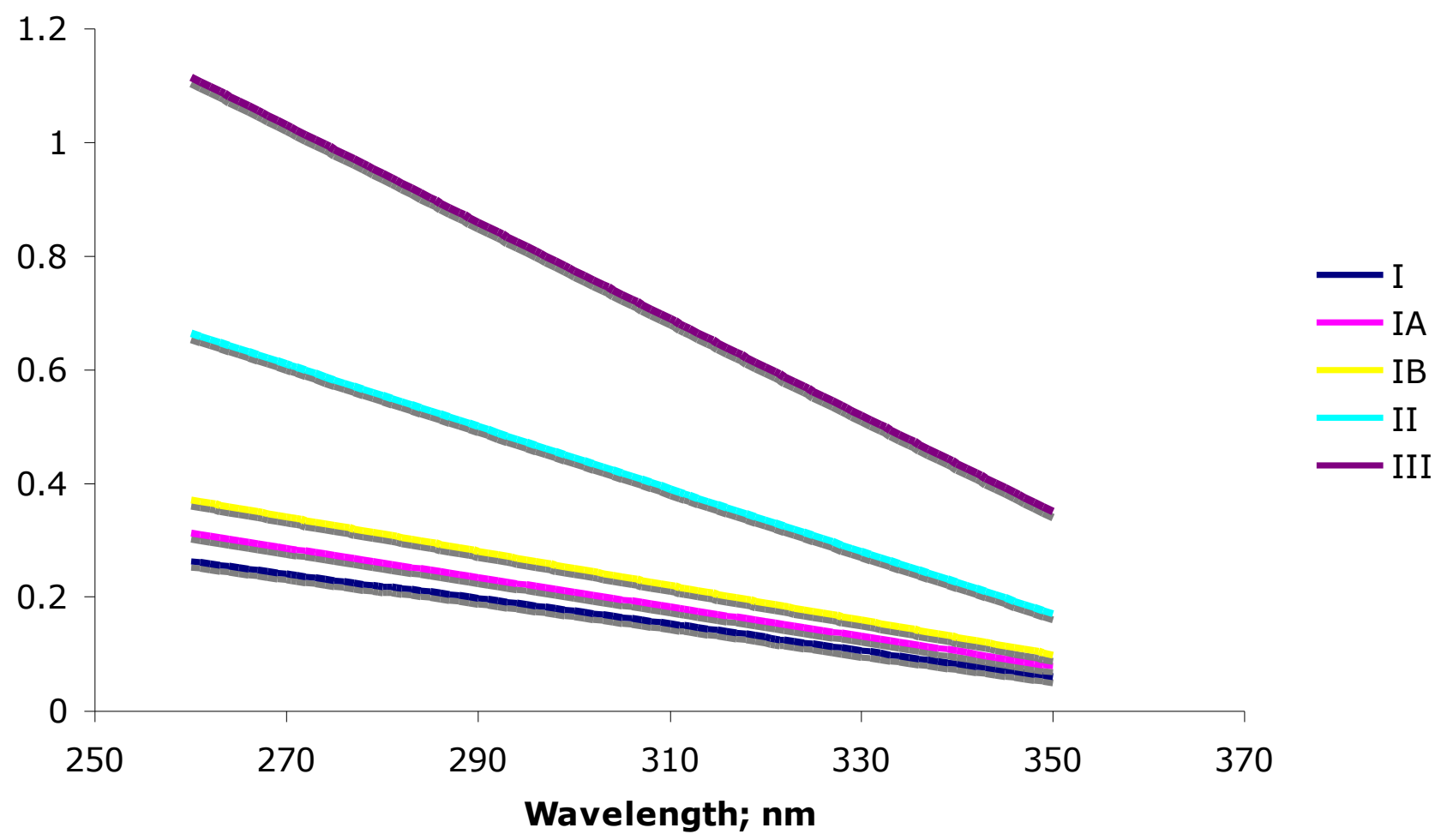

**Fig. 1** Attenuation coefficients for the five optical ocean water types in the wavelength range used.

The effective biological doses $F^*$ delivered in above water types are plotted in Fig. 2, as a function of depth $z$. Again waters of types I, IA and IB follow a more or less similar behaviour.

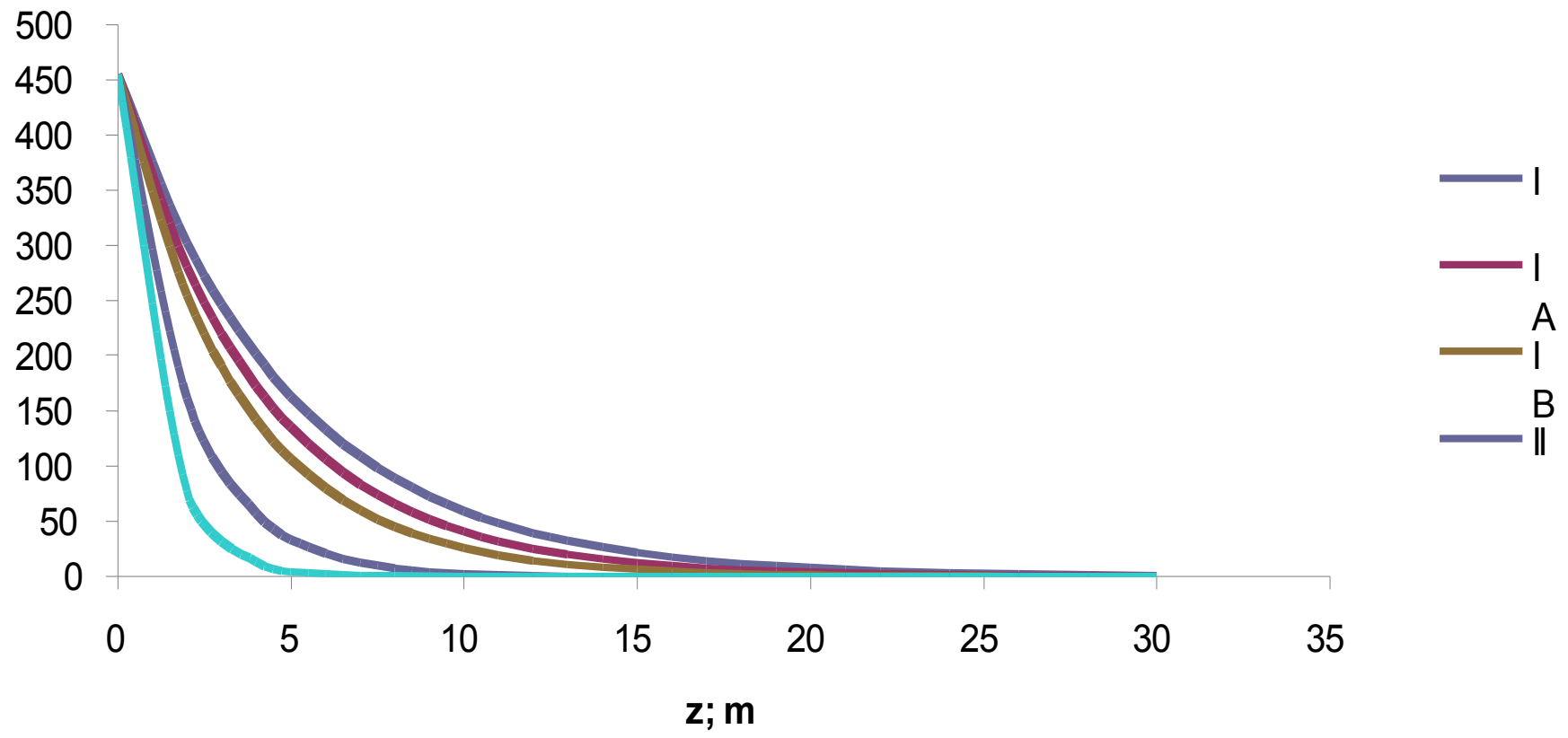

**Fig. 2** Effective biological doses vs. depth for all ocean optical water types

## 3.2 The estimation of induced lethality

In Figs. 3-7 we show the surviving fraction of cells after the GRB UV-flash strikes, for the five optical ocean water types.

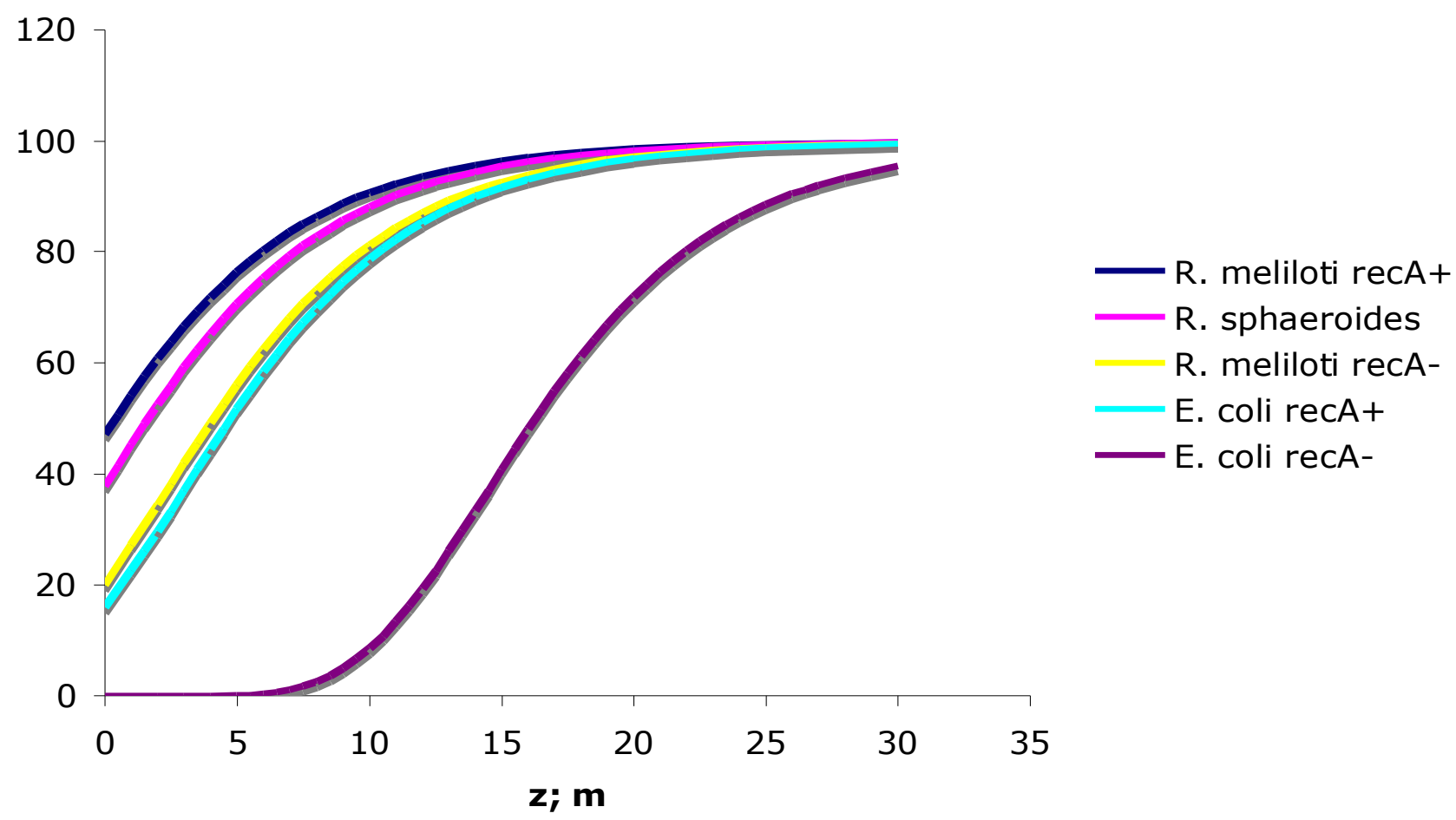

**Fig. 3** Surviving fraction of cells after the GRB UV-flash strikes, for the case of water type I, the more oligotrophic and clear.

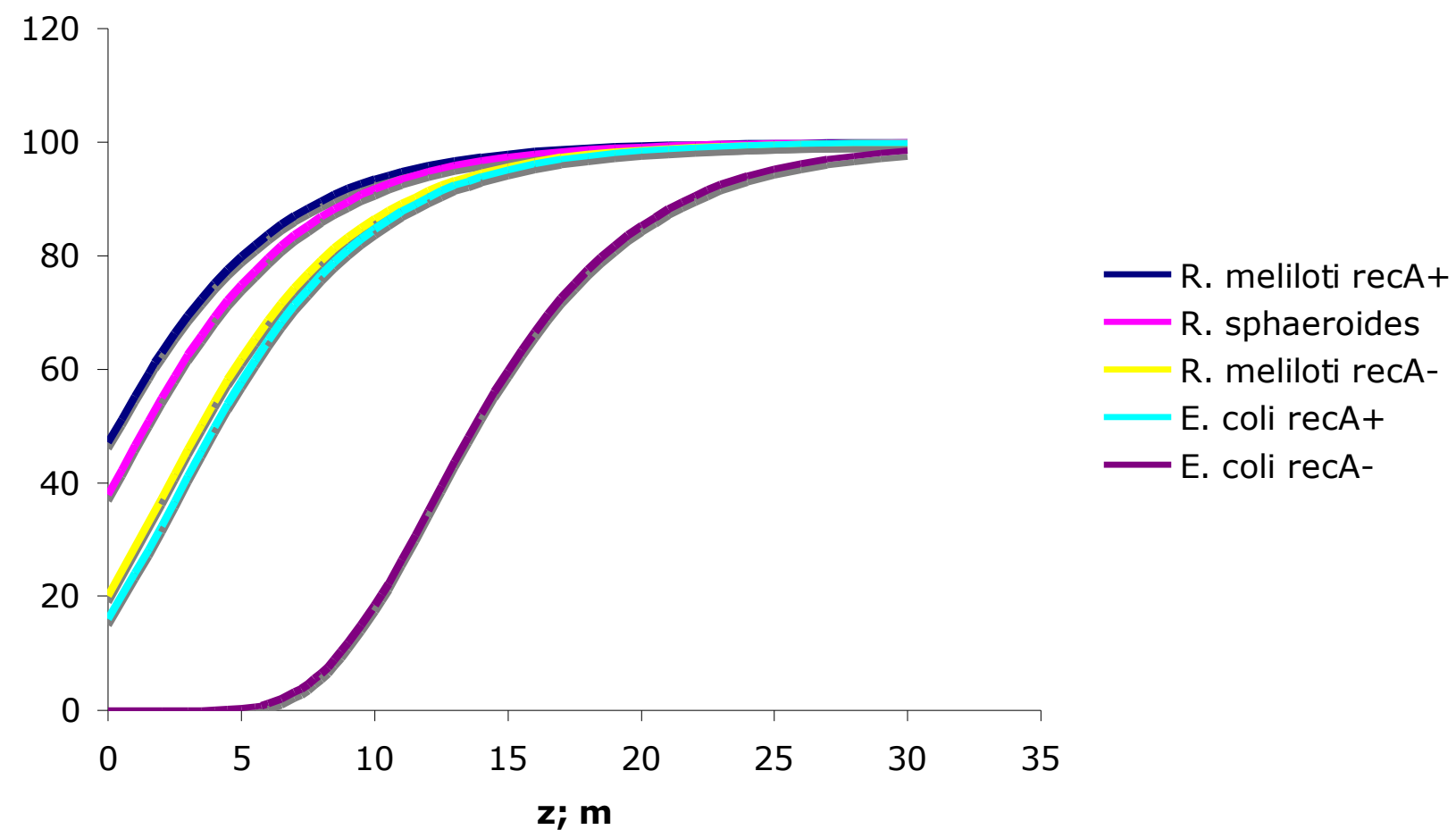

**Fig. 4** Surviving fraction of cells after the GRB UV-flash strikes, for the case of water type IA.

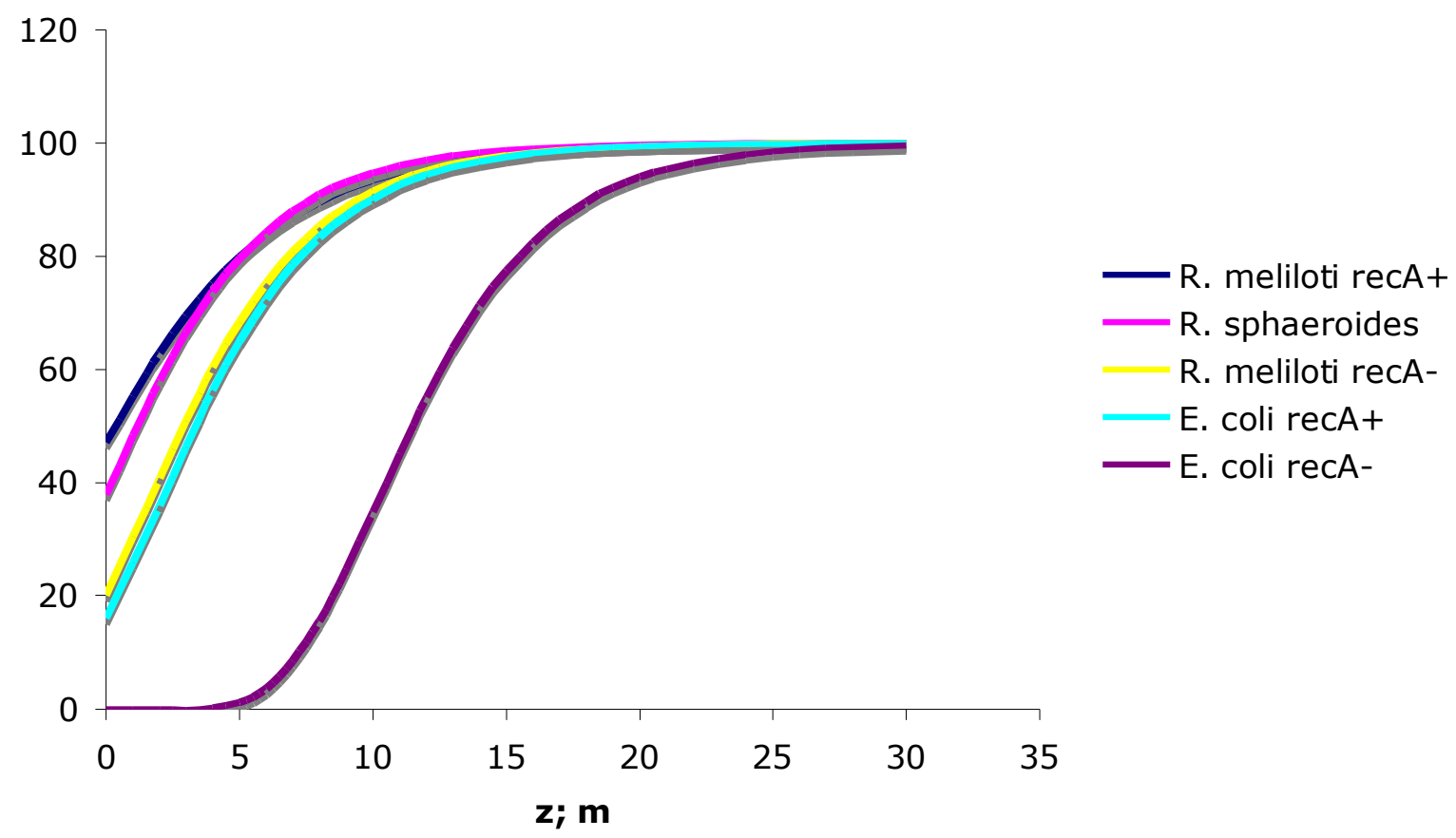

**Fig. 5** Surviving fraction of cells after the GRB UV-flash strikes, for the case of water type IB.

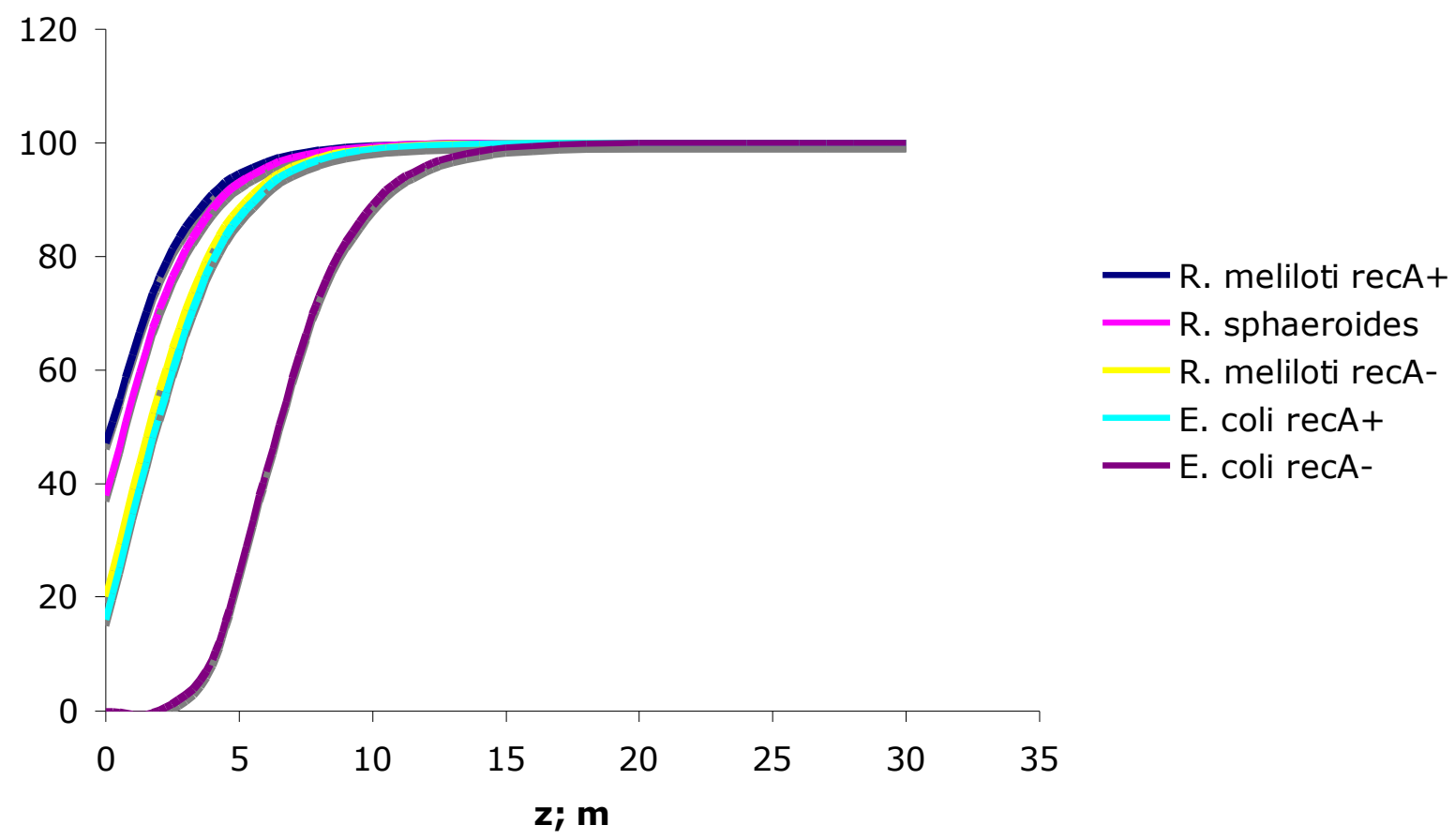

**Fig. 6** Surviving fraction of cells after the GRB UV-flash strikes, for the case of water type II.

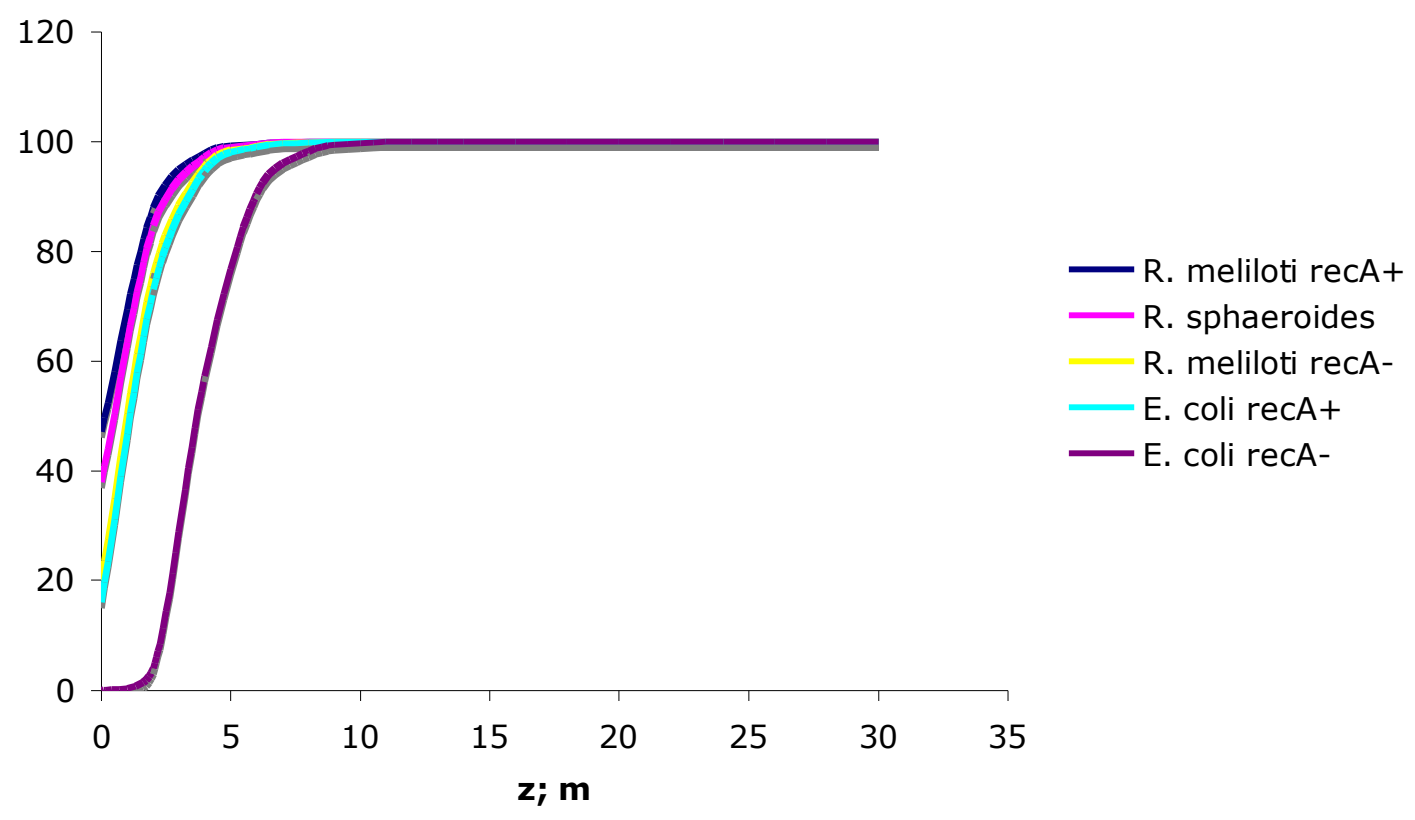

**Fig. 7** Surviving fraction of cells after the GRB UV-flash strikes, for the case of water type III.

In the Table 1 below we present the surviving fractions in the upper mixed layer of the ocean (30 meters depth) resulting from the above calculations.

| | | Total biomass reduction (%) in the mixed layer (30 meters depth) for oceanic optical water types | | | | |
|---|---|---|---|---|---|---|
| Scenario | Species | I | IA | IB | II | III |
| Good repair | *R. meliloti recA*$^+$ | 10,4 | 8,8 | 8,8 | 4,2 | 2,7 |
| | *R. sphaeroides* | 12,8 | 10,8 | 9,1 | 5,2 | 3,3 |
| | *E. coli recA*$^+$ | 20,3 | 17,2 | 14,5 | 8,2 | 5,0 |
| Bad repair (cold water or night UV-flash) | *R. meliloti recA*$^-$ | 18,6 | 15,8 | 13,2 | 7,5 | 4,6 |
| | *E. coli recA*$^-$ | 57,1 | 48,6 | 40,7 | 22,8 | 13,3 |

**Table 1** Total reduction of biomass in the upper mixed layer of the ocean

### 3.3 The role of UV-C

An interesting feature of the action of the GRB is the presence of irradiances at ground level in the very deleterious UV-C band. In this case we computed non-negligible values between 260 and 280nm. This band is not usually considered because it is totally absorbed by the atmosphere, but the aurora-like spectrum provoked by the GRB includes these wavelengths, emitted in the atmosphere at altitudes low enough as to reach ground. Despite its relatively low intensity, this band *per se* would account for a significant lethality, as shown in Table 2.

| | | Total biomass reduction (%) in the mixed layer (30 meters depth) for oceanic optical water types | | | | |
|---|---|---|---|---|---|---|
| Scenario | Species | I | IA | IB | II | III |
| Good repair | *R. meliloti recA*$^+$ | 2,0 | 1,7 | 1,4 | 0,9 | 0,6 |
| | *R. sphaeroides* | 2,6 | 2,2 | 1,8 | 1,1 | 0,8 |
| | *E. coli recA*$^+$ | 4,7 | 3,9 | 3,3 | 2,0 | 1,3 |
| Bad repair (cold water or night UV-flash) | *R. meliloti recA*$^-$ | 4,2 | 3,5 | 3,0 | 1,8 | 1,2 |
| | *E. coli recA*$^-$ | 27,0 | 22,7 | 19,1 | 10,7 | 6,4 |

**Table 1** Total reduction of biomass in the upper mixed layer of the ocean had the UV-flash contained only the UV-C band

## 4 Conclusions

Most areas of modern ocean basins are oligotrophic (water types I, IA and IB), thus from Table 1 we might expect a lethality of ~10% from a gamma-ray illumination in good repair scenarios, assuming that most species of phytoplankton would behave similarly to the aquatic bacterium *R. sphaeroides.* However, the cells of some species of picoplankton are so small (diameter around 0,6 μm), that it is unlikely that they can host an elaborated DNA repair machinery. An outstanding example is the genus *Prochlorococcus*. Due to its wide distribution and small size, *Prochlorococcus spp.* have been termed the most abundant organisms on Earth (Partensky, Hess and Vaulot 1999). In fact, they account for an estimated 20% of the oxygen released to the Earth's atmosphere through the photosynthesis process, and are at the very base of the ocean food assemblage. Given their poor repair capabilities, lethality of species of this genus could reach 25% even in warm waters, and higher at night or in cold waters, where the repair mechanisms are additionally inhibited (Table 1). Also, a night flash would affect several organisms commonly found in deep waters during daylight time. However, it is true that the brief flash of ~ ten seconds would only influence one hemisphere of the planet, the one facing the gamma beam, a fact that attenuates the size of the damage. It is also important to note that phytoplankton living beneath the mixed layer at the moment of the UV-flash would not be affected, even in oligotrophic waters I, IA and IB (Figs. 3-7). In Peñate *et al* (2010) a total (100%) inhibition of photosynthesis down to 75 meters in water type I was estimated. Comparison with Table 1 then leads us to state that this is mainly due to damages of the photosynthetic apparatus, and not to the other possible cause (induced lethality due to DNA damage). Therefore, estimation of the velocity of recovery to normal population numbers becomes very complicated, depending on: the repair mechanisms of both the photosynthetic apparatus and DNA, the ocean circulation bringing unaffected individuals from below the mixed layer and from the unaffected hemisphere, the extent of the depletion of the ozone layer (which would last a decade or so), the potential climate change (we refer interested readers to Thomas *et al* 2005 and Galante and Horvath 2007 for a compilation of many potential effects). Aquatic food webs having a strong dependence on phytoplankton might be very affected and it turns out interesting to evaluate the response of the other primary producers (macroalgae and seagrasses). The darker the water the lesser the affectation, therefore in shallow waters the more protected ecosystems would be some very eutrophic coastal and inland waters.